\documentclass[prl,twocolumn,superscriptaddress,floatfix]{revtex4-1}

\usepackage{amssymb}
\usepackage{graphics,graphicx,color}
\usepackage{dcolumn,bm}
\usepackage{tikz}
\usepackage{float}

\usepackage{epsfig}

\begin{document}


\title{Universal mechanical instabilities in the energy landscape of amorphous solids: evidence from athermal quasistatic expansion}
\author{Umang A. Dattani}
\affiliation{The Institute of Mathematical Sciences, CIT Campus, Taramani, Chennai 600113, India}
\affiliation{Homi Bhabha National Institute, Anushaktinagar, Mumbai 400094, India}
\author{Smarajit Karmakar}
\affiliation{Tata Institute of Fundamental Research, 
36/P, Gopanpally Village, Serilingampally Mandal,Ranga Reddy District, 
Hyderabad, 500107, Telangana, India}
\author{Pinaki Chaudhuri}
\affiliation{The Institute of Mathematical Sciences, CIT Campus, Taramani, Chennai 600113, India}
\affiliation{Homi Bhabha National Institute, Anushaktinagar, Mumbai 400094, India}


\begin{abstract}
Using numerical simulations, We study the failure of an amorphous solid under quasi-static expansion starting from a homogeneous high-density state. During the volume expansion, we demonstrate the existence of instabilities manifesting via saddle-node bifurcation in which a minimum meets a saddle. During all such events, the smallest eigenvalue of the Hessian matrix vanishes as a square-root singularity. The plastic instabilities are manifested via sudden jumps in pressure and energy, with the largest event happening when a cavity appears, leading to the yielding of the material.  We show that during cavitation and prior to complete fracture, the statistics of pressure or energy jumps corresponding to the plastic events show sub-extensive finite-size scaling, similar to the case of simple shear but with different exponents. Thus, overall, our study reveals universality in the fundamental characteristics during mechanical failure in amorphous solids under any quasi-static deformation protocol. 
\end{abstract}

\maketitle


{\em Introduction.} Amorphous solids are ubiquitous, and extensively utilised in diverse applications that span from nanometric scales to centimeters \cite{zallen2008physics, rodney2011modeling, nicolas2018deformation}. Stability of these materials under various external mechanical perturbations, across different scales, is important for use in applications. Therefore, understanding the underlying processes leading to their mechanical failure under various deformation is of great significance for future uses. Cavitation is one such mode of catastrophic failure that occurs in these solids, be it  hard (metallic glass, polymer glass etc.) or soft (emulsions, wet granular matter, gels etc.), when the material is subjected to expansion of some kind. The initial micro-cavities potentially lead to eventual fracture, if the expansion process is continued \cite{jiang2014cryogenic, Bouchaud_2008, doi:10.1063/1.4901281, PhysRevLett.98.235501, PhysRevLett.90.075504, Guerra390, PhysRevLett.95.135501, PhysRevLett.107.215501, PhysRevLett.117.044302, aj1983fracture, mott1993atomistic, rottler2003growth}. Thermodynamically, formation of cavities is related to accessing the zone of solid-gas coexistence in the temperature-density plane of the phase-diagram of cohesive glass-forming systems \cite{Pablo, Pablo2, TestardBerthierKob}. Recently,  the interplay between the external mechanical drive and the ambient thermal fluctuations that lead to cavitation in glass-forming systems has been explored \cite{Falk, Pinaki, CavitySmarajit}. Cavitation can also occur in athermal conditions, when the thermodynamic phase-coexistence boundary is encountered during the mechanical expansion \cite{Pablo}.

The underlying energy landscape of amorphous materials is complex, and analysing  properties of these materials in terms of the landscape's features, especially the local minima (called inherent structures (IS)) and saddles, is a domain of active research. In the context of the material's response to athermal quasistatic shear (AQS) \cite{Lemaitre, barrat2011heterogeneities}, i.e. at vanishing driving rate, it has been demonstrated that during the straining process, yielding happens via the occurrence of irreversible plastic events and each such plastic event  can be considered as a catastrophic process corresponding to a saddle node bifurcation within the energy landscape. This corresponds to the smallest non-zero eigenvalue of the Hessian matrix, $\mathcal{H}_{ij}^{\alpha\beta}$ (defined later), going through zero, whenever a plastic event occurs, and the vanishing of the eigenvalue follows a square root singularity \cite{malandro1999relationships, maloney2004subextensive, Lemaitre, SmarajitElasticConstants, SmarajitYielding}. Further, it has been shown \cite{SmarajitYielding} that, during the steady state under athermal quasi-static shear, the mean drops in energy and stress respectively scale as $\langle\Delta{U}\rangle \sim N^{\alpha} $, $\langle\Delta{\sigma}\rangle \sim N^{\beta}$, where $\alpha=1/3$ and $\beta=-2/3$ being universal exponents in both two and three dimensions  across different model systems. 

It is pertinent to ask whether other quasi-static mechanical deformations also involve similar singularities and follow similar statistics. In the context of cavitation, a recent study using a one-component system, which would potentially crystallize, seemed to show that at the instant of cavitation of the solid,  a saddle-node bifurcation occurs \cite{ShimadaOyama}. However, the study specifically focused near the Sastry density, i.e. where IS states start to show spatial inhomogeneities \cite{PhysRevE.56.5533, PhysRevLett.85.590}, by probing the stability of these states obtained through thermal quench from a high-temperature liquid state. The behaviour at higher or lower densities was not studied. 

{\em Summary.} In this letter, we investigate the stability of local minima within the energy landscape visited by a model amorphous solid, while it is subjected to athermal quasi-static expansion (AQE), starting from a homogeneous high density state. We observe that the expansion leads to the release of pressure via formation of cavities, which become precursors for fracture progression upon continued expansion and  eventually leads to loss of material integrity. We demonstrate that, all throughout this trajectory, plastic events of different magnitudes happen, be it in the homogeneous state or when density inhomogeneities pop-up and proliferate. And, during all such events, the  smallest eigenvalue of the Hessian matrix vanishes as a square root singularity, of the kind discussed above. Further, we show that during cavitation and prior to complete fracture, the statistics of pressure or energy jumps corresponding to the plastic events show sub-extensive finite size scaling, similar to the case of AQS, but with different exponents. Thus, broadly speaking, our observations in the context of response to expansion, in conjunction with previous observation on response to shear deformations, indicates that the nature of mechanical instabilities that the underlying landscape of amorphous materials undergo, during any protocol of  quasi-static deformation, seems to be universal in nature.  

{\em Model and Methods.} For our study, we consider a well-characterized two-dimensional model binary Lennard-Jones mixture, whose mechanical response have been extensively investigated; see SI for details of the model \cite{si}. We study the athermal response of this system to isotropic expansion via a quasi-static  process \cite{Pablo}, starting from a high density state ($\rho=1.2$) where the system is spatially homogeneous having a  positive barostatic pressure. In each expansion step, a constant volume strain is applied on the system by rescaling the length of the box by a factor $(1+\epsilon)$ along with affine transformation of particle coordinates, followed by minimization of the energy of this strained configuration using the conjugate gradient algorithm~\cite{cg}.  The values of $\epsilon$ are varied from $\epsilon=10^{-4}$ to $\epsilon=10^{-9}$. The initial states used for the expansion process are obtained by thermally cooling high temperature liquid states, followed by energy minimization; see SI for further details. The AQE simulations are done using LAMMPS\cite{LAMMPS}.  LAPACKE\cite{lapack99} is used for doing the stability analysis of the local minima states, by computing eigenvalues and eigenvectors of the Hessian matrix $\mathcal{H}^{\alpha \beta}_{i j}$, which is defined as 
\begin{equation}
    \mathcal{H}_{ij}^{\alpha \beta} = \frac{\partial^{2} U
    \left(\{\mathbf{r}_{i}\}\right)}{\partial r_{i}^{\alpha} \partial r_{j}^{\beta}},
    \label{eq_hessian}
\end{equation}
where $U\left(\{\mathbf{r}_{i}\}\right)$ is the potential energy of the system and $\mathbf{r}_i$ is the position vector of particle $i$. The indices $\alpha, \beta \in \{x,y\}$ runs over all dimensions whereas $i,j \in \{1, \ldots , N\}$. Our studies are done for a variety of system sizes ranging from $N=10^3$ to $N=10^5$.

\begin{figure}
    \centering
    \includegraphics[width=0.98\linewidth]{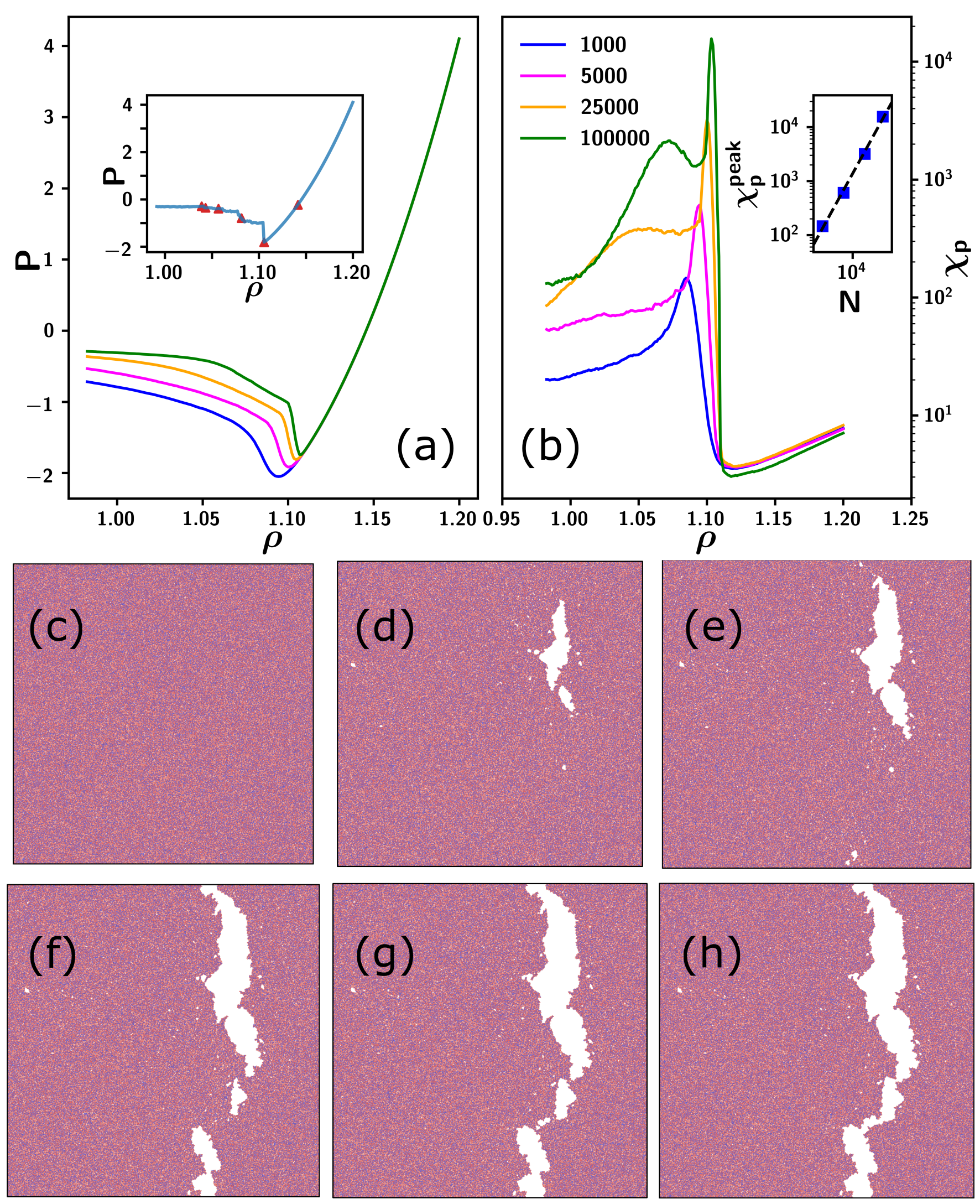}
    \caption{(a) Variation of average pressure ($P$) and (b) corresponding susceptibility ($\chi_p$) with density ($\rho$), for different system sizes as marked. (c)-(h) Snapshots from a $N=10^5$ system showing cavitation and eventual failure, at the density points marked for the single trajectory shown in the inset of (a). Inset of (b) shows the finite size dependence of susceptibility peak, $\chi_p^{peak} \sim N$, indicated with the dashed line.}
        \label{fig1}
\end{figure}

{\em The yielding.} When we expand the homogeneous solid, the density decreases and expectedly the pressure also decreases; see Fig.\ref{fig1}(a) for the density dependence of the ensemble-averaged pressure $\langle{P}\rangle$ for different system sizes. Energy of the system also decreases simultaneously; see SI \cite{si}.  As expansion continues, beyond some density, the pressure becomes negative implying occurrence of internal tension. When the pressure changes sign, the energy goes through a minimum and then starts increasing. Eventually, the built-up tension is released, and the system yields with a big jump in pressure (see inset of Fig.\ref{fig1}(a) for the case of a trajectory corresponding to $N=10^5$), as well as energy (see SI \cite{si}). The fluctuations in pressure within the ensemble, quantified via a susceptibility $\chi_p = N(\langle P^2 \rangle - \langle P \rangle^2)$, goes through a maximum when the jump in pressure is witnessed; see  Fig.\ref{fig1}(b). The pressure jump and a peak in the corresponding susceptibility parallels the response to shear, where a large stress drop at yielding is associated with a peak in stress fluctuations \cite{ozawa2018random}. 

We note that the jump in pressure becomes sharper with increasing system size. Along with that, the peak height of the susceptibility increases as ${\chi_p}^{\rm peak} \sim N$ (see inset of Fig.\ref{fig1}(b)), with the peak also becoming narrower, similar to what has been observed in the case of AQS response \cite{ozawa2018random}. These observations evidence the existence of a yielding transition, located at the density at which ${\chi_p}^{\rm peak}$ occurs, with the yield point shifting to larger densities with increasing $N$. 

We now focus on the spatial ramifications of the response to the AQE process by illustrating the evolution of an example particle configuration, whose $P$ vs $\rho$ trajectory is shown in inset of Fig.\ref{fig1}(a). We observe that the state which is spatially homogeneous at higher density (see Fig.\ref{fig1}(c)), yields under expansion via the formation of a large cavity, as is shown in Fig.\ref{fig1}(d). Upon further expansion, the cavity increases in size and newer cavities appear in front of the expanding front (see Fig.\ref{fig1}(e),(f)), with these events showing up as saw-toothed steps in the $P$ vs $\rho$  as shown in the inset of Fig.\ref{fig1}(a) \cite{Pablo}. While the expansion proceeds, cavities start merging, the fracture expands and eventually percolates leading to the complete failure of the system \cite{bouchbinder2008dynamic, bouchbinder2008stability}; see Fig.\ref{fig1}(g),(h). 

When sampled over the ensemble of independent initial states and corresponding expansion trajectories, the final fracturing event is marked by a secondary peak in $\chi_p$, very distinctively visible for the larger system sizes ($N=10^5, 2.5\times{10^4}$). Here too, we notice that the location of the peak shifts to larger density with increasing system size \cite{si}. We note that this fracture process also happens in the smaller system and the second peak in $\chi_p$ is likely to occur there too, probably at smaller $\rho$.


\begin{figure}
\centering
\includegraphics[width=0.93\linewidth]{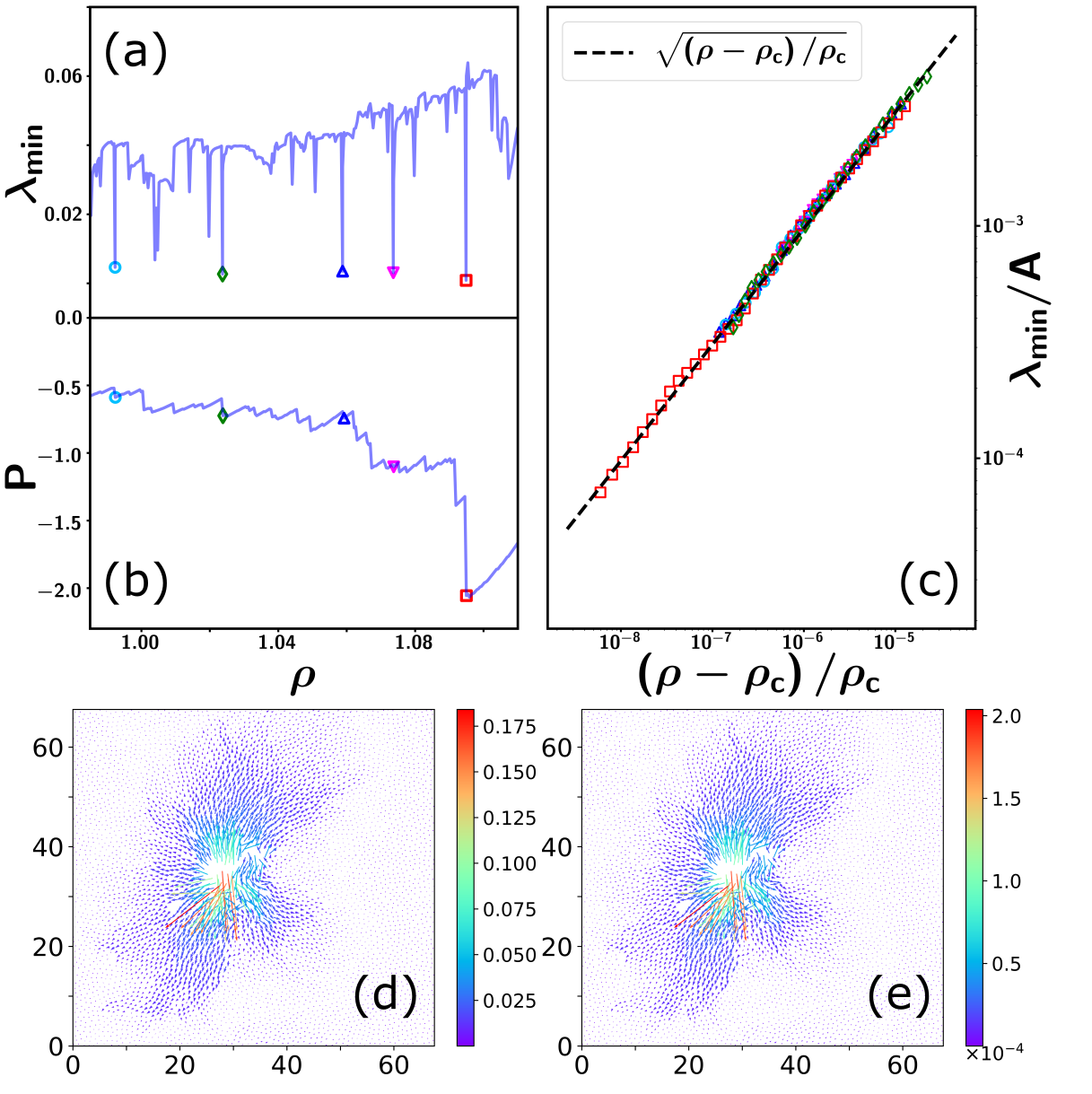}
\caption{(a) Variation of the minimum eigenvalue of the Hessians, $\lambda_{min}$, during a particular quasistatic expansion trajectory of a $N=5000$ system. (b) Corresponding variation of pressure with density. (c) For the amorphous states obtained at the density points marked in (a) and (b), the minimum eigenvalue displays the following power law $\lambda_{min}\propto{\sqrt{(\rho-\rho_c)/\rho_c}}$, where $\rho_c$ is the estimated point of singularity.  For the trajectory shown in (b), at the $\rho_c$ where the largest pressure jump occurs,  (d) the spatial map of the eigenfunction corresponding to $\lambda_{min}$, and (e) the measured non-affine displacement field across the pressure jump.}
\label{fig2}
\end{figure}


{\em Post-yield analysis.} Next, we probe the stability of the local minima visited by the amorphous system during the quasi-static expansion. As discussed in the introduction, for this purpose, we compute the smallest non-zero eigenvalue, $\lambda_{min}$, of the Hessian matrix of the configuration obtained after each combined step of expansion and minimization.  We focus on the density regime around yielding and beyond, prior to the complete fracture failure. We show in Fig.\ref{fig2}(a), how $\lambda_{min}$ evolves with $\rho$ in the case of a $N=5000$ system, using an expansion step of $\epsilon=5\times{10^{-4}}$, for the  expansion trajectory shown in Fig.\ref{fig2}(b) which had been initiated at $\rho=1.2$. We note that for every jump in pressure, $\lambda_{min}$ drops sharply. When well-resolved via smaller expansion steps ranging between $10^{-8}$ to $10^{-11}$, we observe that $\lambda_{min}$ vanishes as a power-law: $\lambda_{min} \sim \sqrt{(\rho-\rho_c)/{\rho_c}}$, where $\rho_c$ is the estimated location of the square-root singularity in each case. In Fig.\ref{fig2}(c), we show the corresponding power-law behaviour of $\lambda_{min}$ in the vicinity of the density locations marked in Fig.\ref{fig2}(b). The density maps for these density points, shown in SI, illustrate that these locations correspond to the sequence of initial cavitation, subsequent increase in the cavity, and also formation of neighbouring cavities. The vanishing of $\lambda_{min}$, in each case,  is thus consistent with the scenario of saddle-node bifurcation discussed  in the introduction. Thus, for each of these irreversible events, the nature of the singularity is exactly the same, and this is observed for all system sizes as shown in SI. Further, if we plot the eigenfunction corresponding to the minimum eigenvalue at the last resolved density point at the brink of singularity, we observe that its spatial structure is very similar to the non-affine displacement field that is generated during the subsequent pressure jump for the irreversible event that follows. In  Fig.\ref{fig2}(d),(e), we demonstrate this to be the case for the main cavitation event, i.e. the large pressure jump, implying that the cavitation process happens via the saddle-node bifurcation with just the lowest eigenvalue of the Hessian matrix of the system completely controlling it. Similar maps for the other jumps, where the cavity expands or merges with nearby cavities are shown in SI. Via these maps, we also note that the pressure release in each case, including the cavitation, is a spatially non-local process having avalanche-like character.  To summarise, generically, for all such events in this regime around yielding and prior to complete fracture, plastic events correspond to incidents of saddle node bifurcation within the underlying potential energy landscape, similar to what has been earlier observed during the response to quasi-static shear \cite{malandro1999relationships, Lemaitre, SmarajitYielding}.

\begin{figure}[t]
    \centering
    \includegraphics[width=0.99\linewidth]{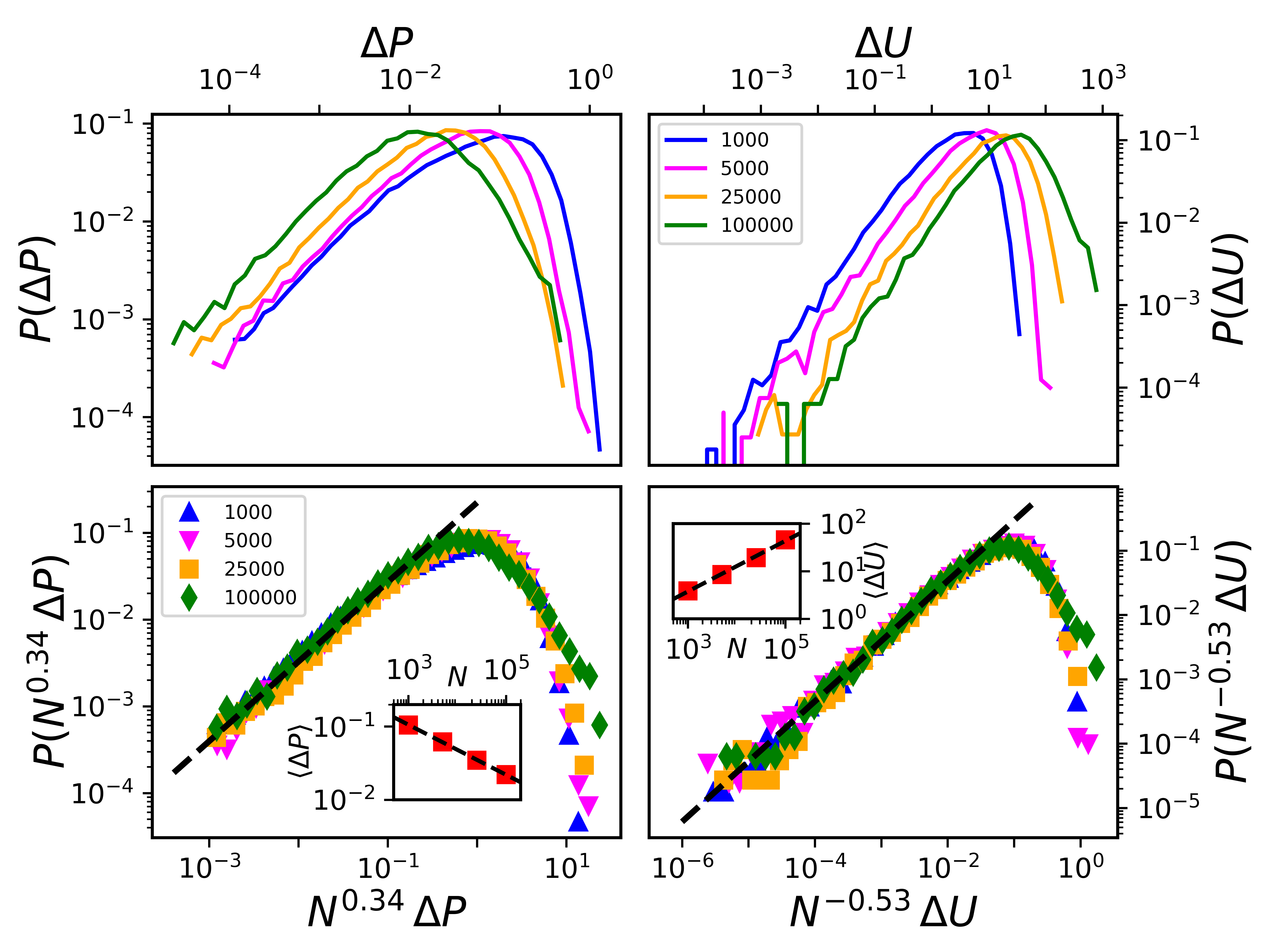}
    \caption{(Top) Distribution of jumps in (left) pressure and (right) energy, during AQE, after occurrence of  main cavitation event. (Bottom) Data collapse of the distributions, with appropriate $N$-scaling, which also reflects in the
    finite size dependence of corresponding mean jump values, viz. $\langle{\Delta{P}}\rangle \sim N^{-0.34}$, $\langle{\Delta{U}}\rangle \sim N^{0.53} $, shown in the respective insets. The dashed lines in bottom panel correspond to power-laws with exponents $0.91$ (left) and $0.93$ (right).}
    \label{fig3}
\end{figure}
 
\begin{figure}[t]
    \centering
    \includegraphics[width=0.9\linewidth]{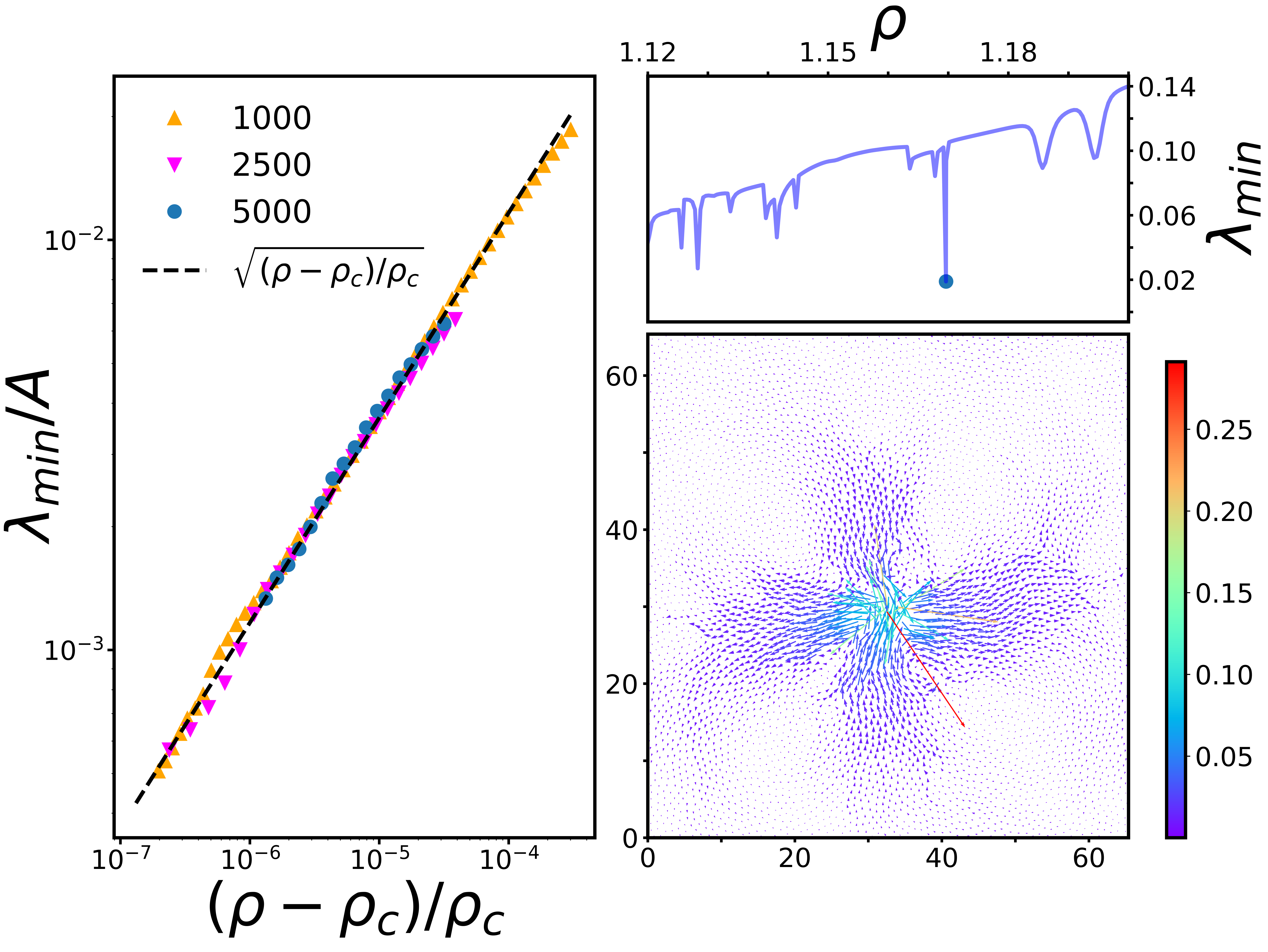}
    \caption{(left) Power law behaviour for the minimum eigenvalue, $\lambda_{min}$, for first events, as a function of $\rho^*$ . Data shown for different system sizes, as marked. (Right) (Top) Variation of $\lambda_{min}$ for a expansion trajectory ($N=5000$), with the location of first event marked. (Bottom) The corresponding map of the eigenfunction related to $\lambda_{min}$ at $\rho \approx \rho_c$. }
    \label{fig4}
\end{figure}

{\em Statistics of plastic events.} We now study the statistics of the jumps in pressure and energy, in the density window between yielding and  complete fracture, for the different system sizes, to probe the finite size scaling behaviour of the distributions of pressure and energy jumps, $\Delta{P}$ and $\Delta{U}$ respectively. We note that in this regime both pressure and energy are not in true stationary state. Nevertheless, we try to get some idea about the scale of the avalanches as the system transits towards eventual failure, after a spot of weakness has been seeded in the form of a cavity. The distributions, $P(\Delta{P})$ and $P(\Delta{U})$, for the different $N$ are shown in Fig.\ref{fig4}(a),(b). The distributions for different $N$ collapse upon scaling the argument by $N^{-0.34}$ and $N^{0.53}$ respectively; see Fig.\ref{fig4}(c),(d), which is also reflected in the $N$-dependence of the first moments of the distributions, viz. $\langle{\Delta{P}}\rangle$ and $\langle{\Delta{U}}\rangle$, as shown in the respective insets. Note that the exponents are different from the case of shear and also the difference between these exponents comes out to be $\approx 0.9$ unlike in the case of yielding under shear where $\alpha-\beta \approx 1$. However, the latter holds only when the density does not change \cite{ItammarArgument}, which is the case for shear but not for expansion. Moreover, the distributions themselves show nice power-law behaviour at small arguments as $P(\Delta{P}) \sim \Delta{P}^{\eta}$ and $P(\Delta{U}) \sim \Delta{U}^{\theta}$, with $\eta \simeq 0.91$ and $\theta \simeq 0.94$
as shown in Fig.\ref{fig4}(c),(d) by the solid lines. It also seems that the forms of the distributions can be approximated by the Weibull distribution with similar power-law exponents; see SI \cite{si}. It is not immediately obvious why pressure and energy drops in the post-yield should show extreme value statistics which is expected for the first plastic drop as shown before for AQS studies. The power-law exponent of $P(\Delta{U})$ suggests a exponent $0.52$ for the system size dependence of $\langle\Delta{U}\rangle$ in agreement with our observation but the pressure drop statistics does not follow the same. This observation does warrant further studies for better understanding plastic instabilities under AQE protocol. Nevertheless, the scaling collapse of distributions suggests that under AQE, post-yield, amorphous solids display scale-free plasticity too just like in the case of AQS albeit with different value of exponents $\alpha$ and $\beta$.    

{\em Analysis of the first event.}  So far, we have been focusing on the events after yielding, i.e. the main cavitation process. Now, we turn our attention to the first instance when $\lambda_{min}$ vanishes, i.e. occurrence of irreversible plastic event, during the expansion process, starting from the high density state. We illustrate the situation in Fig.\ref{fig4}(a), for a single trajectory of a $N=5000$ system, where we observe that such an instability occurs at a fairly large density, far away from cavitation. By sampling such events from several trajectories, we show in Fig.\ref{fig4}(b) that here too $\lambda_{min} \sim \sqrt{(\rho-\rho_c)/{\rho_c}}$, underlying that all such instabilities have a similar origin, viz, a saddle-node bifurcation in the underlying energy landscape. The spatial map of the eigenfunction associated with the instability illustrated in Fig.\ref{fig4}(a) is shown in Fig.\ref{fig4}(c). Again it matches well with the non-affine displacement undergone through the instability (see SI \cite{si}), with both showing system-spanning avalanche-like spatial structures having a quadrupolar Eshelby-like shape \cite{Lemaitre, barrat2011heterogeneities}. We note that even in the case of response to shear, pre-yielding avalanches have been reported \cite{Shang86}. Unlike the case at cavitation, no spatial inhomogeneity in density occurs here; see SI \cite{si}.

In fact, in between this first event and yielding via cavitation, numerous plastic events occur, and in each case, the nature of the singularity is the same. The corresponding spatial maps of non-affine displacements initially show Eshelby-like shapes, which somewhat gets distorted as the cavitation regime is approached (see SI \cite{si}), and at cavitation and beyond take a different structure in the vicinity of the cavity, as is visible in Fig.\ref{fig2}(e). 



{\em Conclusions.} To summarise, we have studied how a model two-dimensional spatially homogeneous amorphous solid responds to quasi-static isotropic expansion. As expected, the release of built-up internal stresses leads to yielding transition via cavitation and then this acts as precursor to eventual failure via system-spanning fracture. Both the yielding density as well as the location of complete fracture failure can be identified via susceptibility measurements. The finite-size scaling of $\chi_p$ is exactly the same as that observed for applied shear. Our main focus is in examining the stability of the local minima that the system visits during the expansion process, by monitoring the smallest eigenvalue of the corresponding Hessian matrix. We demonstrate that whenever the eigenvalue goes through zero, all throughout the expansion trajectory, be it for the first event in the high density phase or around and after the yielding via cavitation, it vanishes as a square root singularity, which is characteristic of a saddle-node bifurcation within the underlying energy landscape. Thus, this indicates that the point of yielding, where such instability was recently reported for a mono-component system \cite{ShimadaOyama}, is nothing special vis-a-vis the nature of how the eigenvalue vanishes. Rather, these irreversible processes, which lead to non-affine displacements, seem universal characteristic to plastic events whenever an amorphous solid responds to large mechanical deformations, be it via shear or expansion. However, we note that unlike yielding via shear which can occur in amorphous materials of all kinds, cavitation has a thermodynamic underpinning vis-a-vis the existence of gas-solid coexistence in the phase diagram of attractive glass-formers, and not for repulsive ones. Future studies will probe the occurrence of such instabilities in higher dimensions.  Also, the statistics of the first and the subsequent intermediate events prior to cavitation need to be further studied in details, to connect with similar analysis done for AQS response.

\textit{Acknowledgments:} We thank the HPC facility at IMSc-Chennai for providing computational resources. S.K. would like to acknowledge the support from Swarna Jayanti Fellowship Grants No. DST/SJF/PSA-01/2018-19 and No.  SB/SFJ/2019-20/05. 


\bibliography{apssamp}

\end{document}